\documentstyle[12pt]{article}

\begin{document}

\topmargin 0pt
\oddsidemargin 7mm
\headheight 0pt
\topskip 0mm
\addtolength{\baselineskip}{0.40\baselineskip}

\hfill SOGANG-HEP 199/95

\hfill May, 1995

\vspace{1cm}

\begin{center}
\large{\bf Symplectic Quantization of the CP$^1$ Model \\
           with the Chern-Simons Term}
\end{center}

\vspace{1cm}

\begin{center}
Yong-Wan Kim, Young-Jai Park, and Yongduk Kim \\
{\it Department of Physics and Basic Science Research Institute,\\
     Sogang University, C.P.O. Box 1142, Seoul 100-611}\\
\end{center}

\vspace{2cm}

The symplectic formalism is fully employed to study the gauge-invariant
CP$^1$ model with the Chern-Simons term.
We consistently
accommodate the CP$^1$ constraint at the Lagrangian level according
to this formalism.

\vspace{1cm}

\newpage

\pagestyle{plain}

Since Dirac [1] introduced the consistent quantization method
for constrained theories, there has been great progress on the
subject about the physical, as well as
the mathematical, properties of theories.
Recently, Faddeev and Jackiw (FJ) [2] suggested
the first-order Lagrangian method for constrained Hamiltonian system.
After their work, Barcelos-Neto and Wotzasek (BW) proposed the
symplectic formalism, which is really an improved version of
FJ's method for the case in which the constraints are not
completely eliminated, and
they applied this formalism to several models [3,4].

On the other hand, the CP$^1$ model with the Chern-Simons term [5,6],
which becomes an archetype example of field theory,
was considered by Polyakov, and he found Bose-Fermi statistics
transmutation in the model [7].
Han [8] has recently analyzed this CP$^1$ model by using
the Dirac formalism together with the first-order Lagrangian method.
In this note, we analyze the CP$^1$ model with the Chern-Simons term
by fully using the symplectic formalism [3],
which is algebraically much simpler than the Dirac formalism.

Our starting Lagrangian for the gauge-invariant CP$^1$ model with the
Chern-Simons term [6,8] is given by
\begin{equation}
{\cal L} = \frac{\kappa}{2\pi} \epsilon^{\mu \nu \rho}
            A_{\mu} \partial_{\nu} A_{\rho}
             + (\partial_{\mu} + {\it i} A_{\mu}) z^*_a
             (\partial^{\mu} - {\it i}
             A^{\mu}) z_a ;~~~ a = 1,2
\end{equation}
with the CP$^1$ constraint
\begin{equation}
\Omega^{(0)}_1 = |z_a|^2 - 1 = 0
\end{equation}
where our convention is $\epsilon^{012}=+1$.
The Lagrangian (1) is invariant under the transformations
$z(x) \rightarrow e^{-i\Lambda (x)} z(x)$ and $A_\mu \rightarrow A_\mu
- \partial_\mu \Lambda (x)$ up to a total divergence.
In order to use the advantage of the symplectic formalism [2,3],
which displays the whole set of symmetries in the symplectic two-form,
first let us, for convenience, although it is not necessary,
make the Lagrangian first-ordered by
introducing auxiliary fields, which are their canonical momenta:
\begin{equation}
\pi_a~\equiv~(\partial_0 + {\it i}A_0)z^*_a,~~~~
\pi_a^*~\equiv~(\partial^0 - {\it i}A^0)z_a
\end{equation}
Then, the desired form of the first-ordered Lagrangian,
including the CP$^1$
constraint, is given by
\begin{equation}
{\cal L}^{(0)}~=~ \frac{\kappa}{2 \pi} \epsilon^{mn} A_n \dot{A}_m
                  + \pi_a \dot{z}_a + \pi^*_a \dot{z}^*_a
                  + \Omega^{(0)}_1 \dot{\alpha} - {\cal H}^{(0)},
                   ~~~m, n=1,2 ,
\end{equation}
where the Hamiltonian is
\begin{eqnarray}
{\cal H}^{(0)} &=& \pi_a \pi^*_a + i A_0 (z_a \pi_a - z^*_a \pi^*_a )
            + (\vec{\nabla} - i \vec{A}) z^*_a \cdot
            (\vec{\nabla} + i \vec{A}) z_a
            - \frac{\kappa}{\pi} A_0 \epsilon^{mn} \partial_m A_n
\end{eqnarray}
and $\alpha$ is a Lagrangian multiplier. Note that the canonical sector,
which is the first four parts of the Lagrangian in Eq. (4),
is understood up to a
total time derivative from the usual symplectic conventions.
We have written the superscript to show the iteration properties
of the procedure.

According to the symplectic formalism [2,3],
we have initial sets of symplectic variables and their conjugate
momenta as follows:
\begin{eqnarray}
(\xi^{(0)})^k &=& (A^m, z_a, z^*_a, \pi_a, \pi^*_a, \alpha, A^0), \nonumber \\
(a^{(0)})_k   &=& (\frac{\kappa}{2 \pi}\epsilon^{mn} A^m, \pi_a, \pi^*_a, 0, 0,
0, 0, \Omega^{(0)}_1, 0).
\end{eqnarray}
{}From the definition of the symplectic two-form matrix,
\begin{equation}
f_{ij}(x,y)~=~\frac{\partial a_j(y)}{\partial \xi^i(x)}
         -  \frac{\partial a_i(x)}{\partial \xi^j(y)},
\end{equation}
we have the following singular symplectic matrix:
\begin{eqnarray}
f_{ij}^{(0)}(x,y) ~=~ \left( \begin{array}{ccc}
                       A^{(0)} & B^{(0)} & C^{(0)} (y)\\
                       -{B^{(0)}}^T & 0 & 0 \\
                       -{C^{(0)}}^T (x) & 0 & 0 \\
                  \end{array}
              \right)
  \delta^2 (x-y) \nonumber \\
\end{eqnarray}
where
\begin{eqnarray}
A^{(0)} \!\!=\!\! \left( \begin{array}{cccccc}
         0 &  -\frac{\kappa}{\pi}   &   0  & 0  &  0  &  0  \\
         \frac{\kappa}{\pi} &   0   &   0  & 0  &  0  &  0  \\
         0 &   0   &   0  & 0  & 0 & 0 \\
         0 &   0   &   0  & 0  & 0 & 0 \\
         0 &   0   &   0  & 0  & 0 & 0 \\
         0 &   0   &   0  & 0  & 0 & 0 \\
           \end{array}
    \right)\!,
B^{(0)} \!\!=\!\! \left( \matrix{
         0 &   0  &  0  &  0  \cr
         0 &   0  &  0  &  0  \cr
         -1 &  0  &  0  &  0  \cr
         0 & -1   &  0  &  0  \cr
         0 & 0   & -1  & 0  \cr
         0 & 0 & 0   & -1   \cr
           }
    \right)\!,
C^{(0)} \!\!=\!\! \left( \matrix{
         0 & 0    \cr
         0 & 0    \cr
         z^*_1 (y) & 0     \cr
         z^*_2 (y) & 0     \cr
         z_1 (y) & 0       \cr
         z_2 (y) & 0       \cr
           }
    \right)\!.
\end{eqnarray}
The above singular matrix, $f^{(0)}_{ij}(x,y)$
has a zero eigenvalue and its eigenfunction.
Especially, the eigenfunction is called zero mode,
$({\tilde \nu}^{(0)})^k
= (0, 0, 0, 0, 0, 0, z^*_a v_{11}, z_a v_{11}, v_{11}, v_{12})$
where $v_{11} (x)$ and $v_{12} (x)$ are
independent and arbitrary functions.
Furthermore, the zero mode $({\tilde \nu}^{(0)})^k$ generates two
constraints $\Omega^{(1)}_2$ and $\Omega^{(1)}_3$ such that
\begin{eqnarray}
0 &=&  \int dx~ ({\tilde \nu}^{(0)})^k (x)
       \frac{\partial}{\partial (\xi^{(0)})^k (x)}
       \int dy ~{\cal H}^{(0)} (\xi(y)) \nonumber \\
  &=&  \int dx~ \{ ~v_{11} ( z^*_a \pi^*_a + z_a \pi_a )
       - v_{12} [ i (z^*_a \pi^*_a - z_a \pi_a)
                  + \frac{\kappa}{\pi} \epsilon^{mn}
                  \partial_m A_n  ] \}  \nonumber \\
  &\equiv&  \int dx~[ v_{11} \Omega^{(1)}_3
                       - v_{12} \Omega^{(1)}_2 ].
\end{eqnarray}
This is because of the fact that Hamilton's equation has been written
in the form of the coupled velocities of the symplectic variables,
{\it i.e.}, $f^{(0)}_{ij}{\dot \xi}^{j (0)}
=\frac{\partial H^{(0)}}{\partial \xi^i}$.
Then, the constraints $\Omega^{(1)}_2$ and $\Omega^{(1)}_3$
are incorporated into the Lagrangian to make
the first-iterated Lagrangian [3] as follows:
\begin{equation}
{\cal L}^{(1)} = \left( \frac{\kappa}{2 \pi} \epsilon^{mn} A_n \right)
                 \dot{A}_m
                 + \pi_a \dot{z}_a + \pi^*_a \dot{z}^*_a
                 + \Omega^{(0)}_1 \dot{\alpha} + \Omega^{(1)}_2 \dot{\beta}
                  + \Omega^{(1)}_3 \dot{\gamma} - {\cal H}^{(1)}
\end{equation}
where the first-iterated Hamiltonian is given by
\begin{eqnarray}
{\cal H}^{(1)}(\xi) &=& {\cal H}^{(0)}(\xi)
                       \mid_{ \Omega^{(1)}_2, \Omega^{(1)}_3 =0 } \nonumber \\
                    &=& \pi_a \pi^*_a
                        + (\vec{\nabla} - i \vec{A})
                        z^*_a \cdot (\vec{\nabla} + i \vec{A}) z_a,
\end{eqnarray}
and $\beta$ and $\gamma$ are Lagrange multipliers.
Through this process, we have reduced the original Hamiltonian
using the constraints $\Omega^{(1)}_2$ and $\Omega^{(1)}_3$.

Once again, let us set $(\xi^{(1)})^k$ and $(a^{(1)})_k$ for
first-iterated symplectic variables
and their conjugate momenta:
\begin{eqnarray}
(\xi^{(1)})^k &=& (A^m, z_a, z^*_a, \pi_a,
                  \pi^*_a, \alpha, \beta, \gamma ), \nonumber \\
(a^{(1)})_k   &=& (\frac{\kappa}{2 \pi} \epsilon^{mn} A^n,
                  \pi_a, \pi^*_a, 0, 0, 0, 0, \Omega^{(0)}_1,
                  \Omega^{(1)}_2 , \Omega^{(1)}_3 ),
\end{eqnarray}
respectively. Then, the symplectic two-form matrix is written as
\begin{eqnarray}
 f_{ij}^{(1)}(x,y) ~&=&~
         \left(  \begin{array} {ccc}
                  A^{(0)} & B^{(0)} & C^{(1)} (y) \\
                  -{B^{(0)}}^T & 0 & D^{(1)} (y) \\
                  -{C^{(1)}}^T (x) & -{D^{(1)}}^T (x) & 0 \\
                  \end{array}
         \right)
      \delta^2 (x-y) \nonumber \\
{}~&\equiv&~ F^{(1)}_{ij} (x,y) \delta^2 (x-y).
\end{eqnarray}
where
\begin{eqnarray}
C^{(1)} (y) = \left(  \matrix{
         0 & + \frac{\kappa}{\pi} \partial^y_2 & 0 \cr
         0 & - \frac{\kappa}{\pi} \partial^y_1 & 0   \cr
         z^*_1 (y) & - i \pi_1 (y) & \pi_1 (y) \cr
         z^*_2 (y) & - i \pi_2 (y) & \pi_2 (y)  \cr
         z_1 (y) &  i \pi^*_1 (y) & \pi^*_1 (y)  \cr
         z_2 (y) & i \pi^*_2 (y) & \pi^*_2 (y) \cr
        }
     \right),
D^{(1)} (y) = \left( \matrix{
     0 & - i z_1 (y) & z_1(y)  \cr
     0 & - i z_2 (y) & z_2(y)  \cr
     0 &  i z^*_1 (y) & z^*_1(y)  \cr
     0 &  i z^*_2 (y) & z^*_2(y)  \cr
      }
     \right). \nonumber
\end{eqnarray}
This matrix is still singular, and there is also a zero mode
$({\tilde \nu}^{(1)})^k
= ( - \partial_i v_{12}, i z_a v_{12},\\ - i z^*_a v_{12},
- i \pi_a v_{12}, i \pi^*_a v_{12}, 0, v_{12}, 0 )$
at this stage of iteration.
However, this zero mode does not generate any additional constraint
because it leads to the following trivial identity:
\begin{eqnarray}
\int dx ({\tilde \nu}^{(1)})^k(\xi) \frac{\partial}{\partial (\xi^{(1)})^k(x)}
        \int dy {\cal H}^{(1)}(\xi) = 0.
\end{eqnarray}
This is exactly the result of having a gauge symmetry in the symplectic
formalism. In fact, we can easily check that the first-iterated Lagrangian
in Eq. (11) is invariant up to a total divergence
under the following transformation:
\begin{equation}
\delta (\xi^{(1)})^k = (\tilde{\nu}^{(1)})^k \eta,
\end{equation}
where $\eta$ is only a function of time, or equivalently
\begin{eqnarray}
\delta A^m &=& - \eta \partial_m v_{12},
{}~~~\delta z_a = i \eta v_{12} z_a, \nonumber \\
\delta z_a^* &=& - i \eta v_{12} z_a^*,
{}~~~\delta \pi_a = -i \eta v_{12} \pi_a, \nonumber \\
\delta \pi_a^* &=& i \eta v_{12} \pi_a^*,
{}~~~\delta \alpha = 0, \nonumber \\
\delta \beta &=&  \eta v_{12},
{}~~~\delta \gamma = 0.
\end{eqnarray}

Now, in order to obtain the desired Dirac brackets, we impose the well-known
Coulomb gauge condition, $\nabla \cdot \vec{A} = 0$,
at the Lagrangian level by using the consistent gauge-fixing procedure
in the symplectic formalism [3].
With this constraint, we can directly obtain the gauge-fixed first-order
Lagrangian from the first-iterated Lagrangian in Eq. (10) as follows:
\begin{equation}
{\cal L}^{(2)}_{GF}~=~ \frac{\kappa}{2 \pi} \epsilon^{mn} A_n {\dot A}_m
                   + \pi_a {\dot z}_a + \pi^*_a \dot{z^*}_a
                   + \Omega^{(0)}_1 \dot{\alpha}
                   + \Omega^{(1)}_2 \dot{\beta}
                  + \Omega^{(1)}_3 \dot{\gamma}
                  + \Omega^{(2)}_4 \dot{\sigma} - {\cal H}^{(2)}
\end{equation}
where $\Omega^{(2)}_4 = \nabla \cdot \vec{A}$
and  $\sigma$ is a Lagrange multiplier.
Furthermore,
$H^{(2)}$ is naturally the second-iterated Hamiltonian
\begin{eqnarray}
{\cal H}^{(2)}(\xi) &=& {\cal H}^{(1)}(\xi) \mid_{\Omega^{(2)} =0} \nonumber \\
             &=& \pi_a \pi^*_a + (\vec{\nabla}
                 - i \vec{A}) z^*_a \cdot (\vec{\nabla} + i \vec{A}) z_a.
\end{eqnarray}
Note that this Hamiltonian, which is simply obtained,
is exactly the well-known reduced physical Hamiltonian of the
original CP$^1$ model with the Chern-Simons term,
which may be obtained through several steps with the three definitions
of the canonical, the total, and the reduced Hamiltonian
in the usual Dirac formalism of constrained systems [1].
Furthermore, this model has only four constraints in the symplectic formalism,
while eight constraints are contained in the Dirac method. Therefore,
the symplectic formalism is algebraically much simpler than
that of Dirac.

Now the symplectic procedure is straight-forwardly treated as just
a second-iterated stage with the following
symplectic variables and their conjugate momenta:
\begin{eqnarray}
(\xi^{(2)})^k &=& (A^m, z_a, z^*_a, \pi_a, \pi^*_a, \alpha, \beta, \gamma,
\sigma ), \nonumber \\
(a^{(2)})_k   &=& (\frac{\kappa}{2 \pi} \epsilon^{mn} A^n, \pi_a, \pi^*_a, 0,
0, 0, 0, \Omega^{(0)}_1, \Omega^{(1)}_2 , \Omega^{(1)}_3 ,\Omega^{(2)}_4 ).
\end{eqnarray}
Following the definition of the symplectic matrix, we then find
the second-iterated non-singular symplectic two-form matrix:
\begin{eqnarray}
f_{ij}^{(2)}(x,y) ~=~ \left( \begin{array}{cc}
                             F^{(1)}(x,y) & F^{(2)} (y)  \\
                             -{F^{(2)}}^T (x) & 0 \\
                             \end{array}
                      \right) \delta^2 (x-y)
\end{eqnarray}
where
\begin{eqnarray}
{F^{(2)}}^T (x) = \left( \matrix{ - \partial^x_i & 0 & 0 & 0 & 0 & 0 & 0 & 0 &
0 & 0 & 0\cr } \right). \nonumber
\end{eqnarray}
Then, we have the inverse as follows:
\begin{eqnarray}
 [f_{ij}^{(2)}]^{-1}(x,y) ~=~  \left( \begin{array} {cc}
          G(y) &  I(y)  \\
          -I^T (x) &  J(y) \\
          \end{array}
    \right)
  \delta^2 (x-y)
\end{eqnarray}
where
\begin{eqnarray}
G &=&\! \left( \begin{array}{ccccc}
           0 & - \frac{i \pi}{\kappa} z_b \epsilon^{ij} \frac{\partial_j
}{\nabla^2 } & \frac{i \pi}{\kappa} z^*_b \epsilon^{ij} \frac{\partial_j
}{\nabla^2 } & \frac{i \pi}{\kappa} \pi_b \epsilon^{ij} \frac{\partial_j
}{\nabla^2 } & - \frac{i \pi}{\kappa} \pi^*_b \epsilon^{ij} \frac{\partial_j
}{\nabla^2 } \\
           + \frac{i \pi}{\kappa} z_a \epsilon^{ij} \frac{\partial_i }{\nabla^2
} & 0 & 0 & \delta_{ab} - \frac{1}{2} z_a z^*_b & -\frac{1}{2} z_a z_b \\
           - \frac{i \pi}{\kappa} z^*_a \epsilon^{ij} \frac{\partial_i
}{\nabla^2 } & 0 & 0 & -\frac{1}{2} z^*_a z^*_b & \delta_{ab} - \frac{1}{2}
z^*_a z_b  \\
           - \frac{i \pi}{\kappa} \pi_a \epsilon^{ij} \frac{\partial_i
}{\nabla^2 } & -\delta_{ab} + \frac{1}{2} z^*_a z_b & \frac{1}{2} z^*_a z^*_b
& -\frac{1}{2} (z^*_a \pi_b - z^*_b \pi_a ) & -\frac{1}{2} (z^*_a \pi^*_b - z_b
\pi_a ) \\
           + \frac{i \pi}{\kappa} \pi^*_a \epsilon^{ij} \frac{\partial_i
}{\nabla^2 } & -\frac{1}{2} z_a z_b & -\delta_{ab} + \frac{1}{2} z_a z^*_b &
\frac{1}{2} (z^*_b \pi^*_a - z_a \pi_b ) & -\frac{1}{2} (z_a \pi^*_b - z_b
\pi^*_a )  \\
          \end{array}
    \right),  \nonumber \\
I &=&\! \left( \matrix{
           0 & - \frac{\pi}{\kappa} \epsilon^{ij} \frac{\partial_j }{\nabla^2 }
& 0 & - \frac{\partial_i }{\nabla^2 }  \cr
           -\frac{1}{2} z_a & 0 & 0 & i z_a \frac{1}{\nabla^2}  \cr
           -\frac{1}{2} z^*_a & 0 & 0 & -i z_a \frac{1}{\nabla^2}  \cr
           -\frac{1}{2} \pi_a & 0 & -\frac{1}{2} z^*_a & -i \pi_a
\frac{1}{\nabla^2}  \cr
           -\frac{1}{2} \pi^*_a & 0 & -\frac{1}{2} z_a & -i \pi^*_a
\frac{1}{\nabla^2}  \cr
          }
    \right),
J =\! \left( \begin{array}{cccc}
           0  & 0  & -\frac{1}{2}  & 0 \\
           0  &  0  &  0  & \frac{1}{\nabla^2} \\
           \frac{1}{2}  &  0  &  0  &  0  \\
           0  &  - \frac{1}{\nabla^2}  &  0  &  0 \\
          \end{array}
    \right) . \nonumber
\end{eqnarray}
As results, since $\{ \xi^{(2)i}(x),\xi^{(2)j)}(y) \}
= (f^{(2)})^{-1}_{ij} (x,y)$
according to the FJ method [2,3],
we directly read off the well-known results [6,8]
of the nonvanishing Dirac brackets from this inverse matrix
as follows:
\begin{eqnarray}
& &\left\{ A^m(x),~z_a(y) \right\}_{D} = -\frac{i \pi}{\kappa} z_a
\epsilon^{mn} \frac{\partial^x_n }{\nabla^2 } \delta^2 (x-y), \nonumber \\
& &\left\{ A^m(x),~z^*_a(y) \right\}_{D} = \frac{i \pi}{\kappa} z^*_a
\epsilon^{mn} \frac{\partial^x_n }{\nabla^2 } \delta^2 (x-y), \nonumber \\
& &\left\{ A^m(x),~\pi_a(y) \right\}_{D} = -\frac{i \pi}{\kappa} \pi_a
\epsilon^{mn} \frac{\partial^x_n }{\nabla^2 } \delta^2 (x-y), \nonumber \\
& &\left\{ A^m(x),~\pi^*_a(y) \right\}_{D} = \frac{i \pi}{\kappa} \pi^*_a
\epsilon^{mn} \frac{\partial^x_n }{\nabla^2 } \delta^2 (x-y), \nonumber \\
& &\left\{ z_a(x),~\pi_b(y) \right\}_{D} = (\delta_{ab} -\frac{1}{2} z_a z^*_b)
\delta^2 (x-y), \nonumber \\
& &\left\{ z_a(x),~\pi^*_b(y) \right\}_{D} = -\frac{1}{2} z_a z_b \delta^2
(x-y), \nonumber \\
& &\left\{ z^*_a(x),~\pi_b(y) \right\}_{D} = -\frac{1}{2} z^*_a z^*_b \delta^2
(x-y),\\
& &\left\{ z^*_a(x),~\pi^*_b(y) \right\}_{D} = (\delta_{ab} -\frac{1}{2} z^*_a
z_b)\delta^2 (x-y) , \nonumber \\
& &\left\{ \pi_a(x),~\pi_b(y) \right\}_{D} = -\frac{1}{2} \left( z^*_a \pi_b -
z^*_b \pi_a \right)\delta^2 (x-y), \nonumber \\
& &\left\{ \pi_a(x),~\pi^*_b(y) \right\}_{D} = -\frac{1}{2} \left( z^*_a
\pi^*_b - z_b \pi_a \right)\delta^2 (x-y), \nonumber \\
& &\left\{ \pi^*_a(x),~\pi^*_b(y) \right\}_{D} = -\frac{1}{2} \left( z_a
\pi^*_b - z_b \pi^*_a \right)\delta^2 (x-y). \nonumber
\end{eqnarray}

It seems appropriate to comment on the quantization of the system.
For the simple case that the operator-ordering problem does not exist,
we can directly replace the Dirac brackets with the quantum
commutator: $\left\{~,~\right\}_{D} \rightarrow - i [~,~]$.
However, we should carefully treat the CP$^1$ case
having the ordering problem [6].
For this CP$^1$ model, Han has already found the correct Dirac Brackets [8].
On the other hand, an other effective formalism [9,10]
exists which avoids the ordering problem,
a kind of BFV-BRST method [11]. Recently, this formalism
has been successfully applied to the $CP^{N-1}$ model [12].

In conclusion, we consistently accommodate the CP$^1$ constraint
at the Lagrangian level according to the symplectic formalism.
As a result,
we have explicitly considered the gauge-invariant $CP^1$ model with the
Chern-Simons term by fully using the symplectic formalism,
comparing it with the Dirac method.

\vspace{1cm}

\begin{center}
{\bf ACKNOWLEDGMENTS}
\end{center}

The present study was supported in part by
Sogang University Research Grants in 1995
and by the Basic Science Research
Institute Program, Ministry of Education, Project No. BSRI 95-2414.

\newpage

\begin{center}
{\bf REFERENCES}
\end{center}

\begin{description}
\item{[1]}  P. A. M. Dirac, {\it Lectures on Quantum Mechanics}
            (Belfer Graduate School of Science, Yeshiva Univ. Press,
            New York, 1964).
\item{[2]}  L. Faddeev and R. Jackiw, Phys. Rev. Lett. {\bf 60}, 1692 (1988).
\item{[3]}  M. M. Horta Barreira and C. Wotzasek,
            Phys. Rev. {\bf D 45}, 1440 (1992);
            J. Barcelos-Neto and C. Wotzasek,
            Mod. Phys. Lett. {\bf A 7}, 1737 (1992);
            Int. J. Mod. Phys. {\bf A 7}, 4981 (1992);
            J. W. Jun and C. Jue, Phys. Rev. {\bf D 50}, 2939 (1994);
            Y-W. Kim, Y-J. Park, K. Y. Kim, and Y. Kim,
            J. Korean Phys. Soc. {\bf 26}, 243 (1993);
            S-J. Yoon, Y-W. Kim, S-K. Kim, Y-J. Park, K. Y. Kim, and Y. Kim,
            {\it ibid.}, {\bf 27}, 270 (1994);
            Y-W. Kim, Y-J. Park, K. Y. Kim, and Y. Kim,
            {\it ibid.}, {\bf 27}, 610 (1994);
            S-K. Kim, Y-W. Kim, Y-J. Park, Y. Kim, C-H. Kim, and W. T. Kim,
            {\it ibid.}, {\bf 28}, 128 (1995).
\item{[4]}  E-B. Park, Y-W. Kim, Y-J. Park, Y. Kim, and W. T. Kim,
            Mod. Phys. Lett. {\bf A 10}, 1119 (1995).
\item{[5]}  S. Deser, R. Jackiw and S. Templeton,
            Ann. Phys. (N.Y.) {\bf 140}, 372 (1982).
\item{[6]}  P. K. Panigrahi, Shibaji Roy, and Wolfgang Scherer,
            Phys. Rev. Lett. {\bf 61}, 2827 (1988).
\item{[7]}  A. M. Polyakov, Mod. Phys. Lett. {\bf A 3}, 325 (1988).
\item{[8]}  C. Han, Phys. Rev. {\bf D 47}, 5521 (1993).
\item{[9]}  I. A. Batalin and I. V. Tyutin,
            Int. J. Mod. Phys. {\bf A 6}, 3255 (1991).
\item{[10]}  R. Banerjee, Phys. Rev. {\bf D 48}, R5467 (1993);
            W. T. Kim and Y-J. Park, Phys. Lett. {\bf B336}, 396 (1994);
            Y-W. Kim, Y-J. Park, K. Y. Kim, and Y. Kim,
            Phys. Rev. {\bf D 51}, 2943 (1995);
            J-H. Cha, Y-W. Kim, Y-J. Park, Y. Kim, S-K. Kim, and
            W. T. Kim, Z. Phys. {\bf C}, to be published (1995).
\item{[11]}  C. Becci, A. Rouet, and R. Stora,
            Ann. Phys. (N.Y.) {\bf 98}, 287 (1976);
            I. A. Batalin and E. S. Fradkin,
            Phys. Lett. {\bf B180}, 157 (1986);
            Y-W. Kim, S-K. Kim, Y-J. Park, K. Y. Kim, and Y. Kim,
            Phys. Rev. {\bf D 46}, 4574 (1992).
\item{[12]}  N. Banerjee, S. Ghosh, and R. Banerjee,
            Phys. Rev. {\bf D 49}, 1996 (1994).
\end{description}
\end{document}